\documentclass[a4paper,12pt,preprint]{revtex4}
\usepackage{amsmath,mathrsfs,amscd,graphicx}
\usepackage{color}
\usepackage{cancel}
\usepackage{multirow}
\usepackage{rotating}
\usepackage{slashed}
\usepackage{subfig}
\usepackage{textcomp}
\usepackage[final]{pdfpages}
\usepackage{array}
\usepackage{float}
\usepackage{url}
\usepackage{textcomp}
\usepackage{amsfonts, latexsym, epsfig}
\usepackage{bm}
\usepackage{times}
\usepackage{epsfig}
\usepackage{amssymb}
\usepackage{tikz}
\usepackage{slashed}
\usepackage{hyperref}
\usepackage{verbatim}
\usepackage[normalem]{ulem}
\def\beq{\begin{equation}}
\def\eeq{\end{equation}}
\def\be{\begin{equation}}
\def\ee{\end{equation}}
\def\bea{\begin{eqnarray}}
\def\eea{\end{eqnarray}}

\begin{document}
\title{Collider Phenomenology of $e^{-}e^{-}\to W^{-}W^{-}$}
\author{Kai Wang}
\author{Tao Xu}
\author{Liangliang Zhang}
\affiliation{Zhejiang Institute of Modern Physics and Department of Physics, Zhejiang University, Hangzhou, Zhejiang 310027, China 
}
\begin{abstract}
The Majorana nature of neutrinos is one of the most fundamental questions in  particle physics. It is directly related to the violation of accidental lepton number symmetry. This motivated enormous efforts into the search of such process and among them, one conventional experiment is the neutrinoless double-beta decay ($0\nu\beta\beta$). On the other hand, there have been proposals of future electron-positron colliders as ``Higgs factory" for the precise measurement of Higgs boson properties and it has been proposed to convert such machine into an electron-electron collider. This option enables a new way to probe TeV Majorana neutrino via the inverse $0\nu\beta\beta$ decay process ($e^{-}e^{-}\to W^- W^-$) as an alternative and complementary test to the conventional $0\nu\beta\beta$ decay experiments. In this paper, we investigate the collider search for $e^{-}e^{-}\to W^- W^-$ in different decay channels at future electron colliders. We find the pure hadronic channel, semi-leptonic channel with muon and pure leptonic channel with dimuon have the most discovery potential. 
\end{abstract}

\maketitle
Enormous neutrino oscillation experiments in the last two decades have provided definite evidence for non-zero neutrino masses and the mixing between different flavors~\cite{Bilenky:1987ty,Bilenky:2002aw,GonzalezGarcia:2002dz}. Even though the recent discovery of a Higgs-like boson has significantly improved our knowledge over generation of SM fermion masses, being tiny but electric neutral, the origin of neutrino mass may remain an open question. 
Firs of all, if neutrino masses arise from Yukawa couplings as the same mechanicsm as quarks and charged leptons, one immediately encounters the $\mathcal{O}(10^{-12})$ hierarchy in ${y_{\nu}}/{y_{t}}$. A second argument arises from the prediction of electric charge quantization. 
Anomaly-free conditions determine $U(1)_Y$ as the unique $U(1)$ gauge symmetry in SM up to a normalization factor~\cite{Geng:1988pr}. Though extending SM with milli-charged Dirac neutrino does not explicitly violate the anomaly-free conditions, the hyper-charge assignment is no longer uniquely determined unless the neutrino is a Majorana particle ~\cite{Babu:1989ex}. On the other hand, the bound on neutrino electric charge $Q_{\nu}$ is $|Q_{\nu}|\lesssim(0.5\pm2.9)\times10^{-21}e$ (68\% CL) by assuming charge conservation in $\beta$-decay $n\rightarrow p+e^{-}+\bar{\nu}_{e}$~\cite{Dylla:1973zz,Baumann:1988ue}, and $|Q_{\nu}|<2\times10^{-15}e$ from SN1987A astrophysics observation~\cite{Barbiellini:1987zz}.
These facts motivate the study of Majorana neutrinos.  

Taking the effective theory approach, Majorana mass term is from the non-renormalizable Weinberg operator $(y_{ij}/\Lambda_{\cancel{L}})\ell_{i}\ell_{j}\Phi\Phi$~\cite{Weinberg:1979sa} with dimensionless coupling $y_{ij}$. This dimension-five operator breaks lepton number by two units ($\Delta L=2$) and indicates new physics at some specific $\Lambda_{\cancel{L}}$ scale. One elegant observation is that $\cal {O}$(eV) neutrino mass can be a consequence of $M_{\rm GUT}$ suppression. The simplest realization is the so-called type-I ``seesaw" mechanism where a SM singlet neutrino $N$ forms Dirac mass term $y_\nu\bar{\ell_L}N \Phi$ with leptonic $SU(2)_L$ doublet and a Majorana mass term $M_R \overline{N^c} N$ by itself ~\cite{Minkowski:1977sc, Mohapatra:1979ia, Yanagida:1979as, GellMann:1980vs}. The SM singlet $N$ can be accommodated in the spinor representation of $SO(10)$ GUT representation as $16=10+\bar{5}+1$. The lighter mass eigenstates are then identified as light neutrinos and the heavy ones with mass $M_{N}\sim M_{\rm GUT}$ can only be searched for through indirect effects. 

Further access to low seesaw scales exists in extended models where higher-dimensional Weinberg operator $[\mu^{(n-1)}_{ij}/\Lambda_{\cancel{L}}^{n}]\ell_{i}\ell_{j}\Phi\Phi$ allows more freedom in choosing $\Lambda_{\cancel{L}}$ scale and $\mu$ coefficient for neutrino mass generation. 
The low-scale ``seesaw" extension, on the other hand, calls for heavy neutrino $\nu_N$ searches at various scales. At present, there're several types of such experiments but not a specific one to cover all regions. Among them, the $0\nu\beta\beta$ decay experiments is the most important one to discover the lepton number violating (LNV) process with $\Delta L=2$. So far there's no signal event observed by GERDA and KamLAND-Zen collaborations~\cite{Agostini:2013mzu,Gando:2012zm}. This provides the strongest bounds on the neutrino mixing $|V_{eN}|^2$ below $10^{-8}$$\sim$$10^{-6}$ in a wide $M_{N}$ window from $1$ MeV to $500$ GeV. However, this bound is significantly weakened when there're more than two Majorana neutrino flavors  because Majorana $CP$ phases  introduce cancellation between $0\nu\beta\beta$ decay amplitudes~\cite{Wolfenstein:1981rk,Leung:1984vy}. There are as well direct and indirect constraints when $M_{N}$ varies from eV to TeV\cite{Aaij:2014aba,Liventsev:2013zz,Bergsma:1985is, Vilain:1994vg, Vaitaitis:1999wq, Bernardi:1987ek, Artamonov:2014urb,PIENU:2011aa, Hayano:1982wu, Abreu:1996pa, Aad:2015xaa, Badier:1985wg, Baranov:1992vq, CooperSarkar:1985nh, Gallas:1994xp, Astier:2001ck, Khachatryan:2016olu, Khachatryan:2015gha}. Experiments with abundant mesons could probe light $\nu_N$ in meson decay $X^{\pm}\to\ell^{\pm}\nu_N$. The branching ratio is proportional to $|V_{\ell N}|^{2}$ and the lepton spectrum deviates from the usual active neutrino case. Further detection of decays with same-sign dilepton could be evidence of the Majorana property. In LHCb and BELLE experiments where precise $B$-meson measurement is available, LNV decay constrains on $|V_{\ell N}|^{2}$ is around $\mathcal{O}(10^{-4})$ with $M_N$ close to $m_B$\cite{Aaij:2014aba,Liventsev:2013zz}.  For regions below $m_D$, the dubbed beam dump search could detect decay products of those $\nu_N$ from $D$-mesons. The CHARM and NuTeV experiments could respectively push $|V_{eN}|^2$ and $|V_{\mu{N}}|^2$ to below $10^{-6}$ while the PS191 and E949 bounds below $450$~MeV are even stronger\cite{Bergsma:1985is, Vilain:1994vg, Vaitaitis:1999wq, Bernardi:1987ek, Artamonov:2014urb}. The most severe bound in this region is close to $10^{-9}$ when $M_N$ is around $300$~MeV. For even smaller $M_N$, the $E_{\ell}$ peak strategy could be used, for example, in $\pi\to eN$\cite{PIENU:2011aa} and $K\to \mu N$\cite{Hayano:1982wu} processes. When $\nu_N$ are heavier than mesons, the DELPHI experiment at LEP measured $Z\rightarrow \nu_{N}\nu$ branching ratio for $M_{N}$ between $3.5$ and $50$~GeV and the corresponding $|V_{\ell N}|^2$ bound is at $\mathcal{O}$($10^{-5}$)~\cite{Abreu:1996pa}. As for hadron collider searches, the smoking gun signature is same-sign dilepton plus jets without $\slashed{E}_T$. Both the ATLAS\cite{Aad:2015xaa} and CMS\cite{Khachatryan:2016olu, Khachatryan:2015gha} collaborations have published results with $8$~TeV data for $M_N$ up to $500$~GeV. However, they're still weaker than the electroweak precision observable(EWPO) bound from constraining the non-unitarity of leptonic mixing matrix \cite{Antusch:2014woa}
 \beq
 \sum_{i} |V_{ei}|^2\leq 2.1\times 10^{-3}.
 \label{EWPObound}
 \eeq
 More detailed analyses are available in~\cite{Deppisch:2015qwa, Drewes:2015iva, Atre:2009rg}.

\begin{figure}[H]
\centering
\subfloat[$t$-channel]{\includegraphics[scale=0.55]{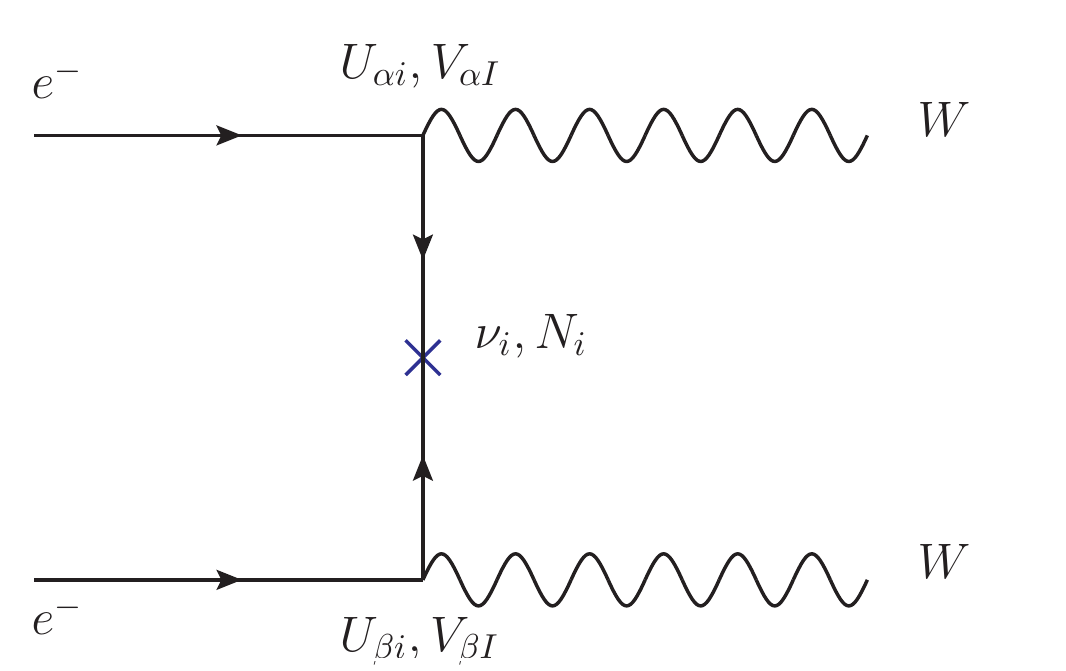}}\quad\subfloat[$u$-channel]{\includegraphics[scale=0.55]{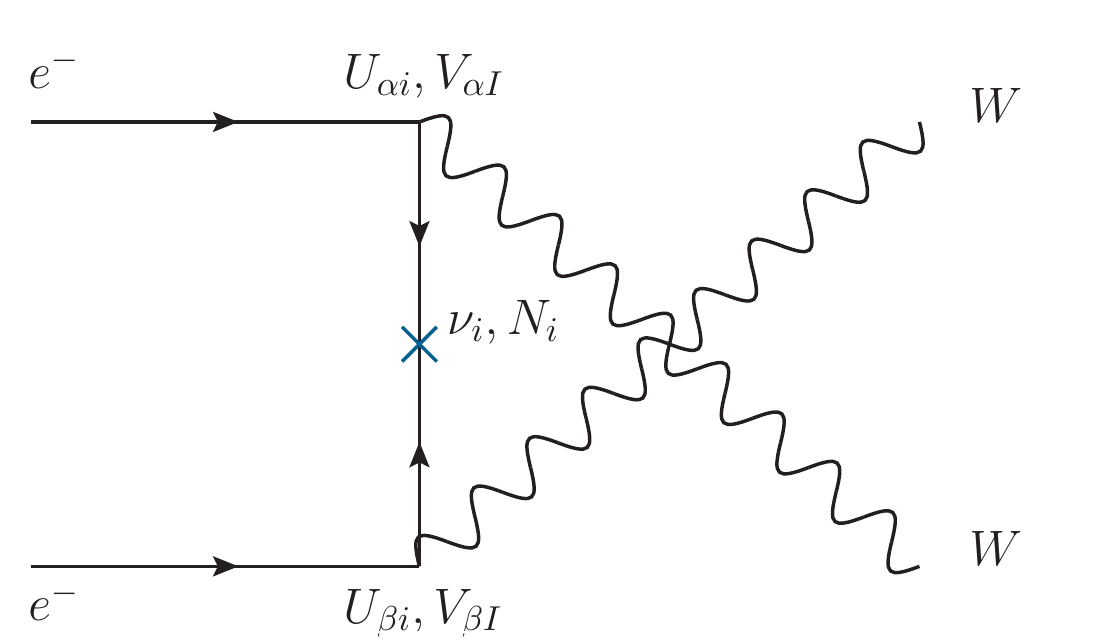}}
\caption{inverse $0\nu\beta\beta$ decay }
\label{eeww}
\end{figure}

As an alternative, $e^{-}e^{-}\to W^{-}W^{-}$ scattering process in Fig.\ref{eeww} mediated by Majorana neutrino exchange is sensitive to the TeV-seesaw scenario. The intriguing feature of this process is that it could be regarded as the inverse of $0\nu\beta\beta$ decay with LNV but could occur at colliders. In addition, the destructive interference effects due to Majorana $CP$ phase in $0\nu\beta\beta$ decay experiments may behave differently as a result of energy scale dependence. According to \cite{Belanger:1995nh}, the unitarity of this process is automatically preserved with the seesaw relation of left-handed electron neutrino Majorana mass. In some extended models with Higgs triplet, this process could also be mediated by a doubly-charged Higgs boson in $s$ channel and this case has been studied in~\cite{Rizzo:1982kn, London:1987nz, Gluza:1995ky, Rodejohann:2010jh}. Previous work on $e^{-}e^{-}\to W^{-}W^{-}$ search could be found in~\cite{Ananthanarayan:1995cn,Rizzo:1982kn, London:1987nz,Dicus:1991fk,Belanger:1995nh, Gluza:1995ky,Gluza:1995ix,Gluza:1995js, Greub:1996ct, Rodejohann:2010jh, Banerjee:2015gca,Asaka:2015oia}. 

Recently, several future electron-positron colliders have been proposed for precise Higgs measurement. Such collider could probe inverse $0\nu\beta\beta$ decay process when converted to an electron-electron machine.
In a most recent study \cite{Asaka:2015oia}, it's shown clearly that the signal cross section could reach  fb level when there're three heavy Majorana neutrinos~($N_{I}$, $I$=1,2,3) with hierarchical masses $M_{1}\ll M_{2}\ll M_{3}$. The mixing $|V_{e2}|^{2}$ of the second heavy Majorana neutrino $N_2$ could be large for $e^{-}e^{-}\to W^{-}W^{-}$ signal production because the $|V_{e1}|^{2}$ and $|V_{e3}|^{2}$ are suppressed by the hierarchical mass relation. In the meantime, the GERDA and KamLAND-Zen constraints could be avoided by destructive interference between $N_1$ and $N_2$. On the other hand, a detailed collider phenomenology study is missing in previous studies and this paper is to fill in the gap by providing studying in all decay channels and focusing on the kinematic methods to reduce background influence on sensitivity. In section II, we discuss the kinematic properties of $e^{-}e^{-}\to W^{-}W^{-}$ process and how to reduce background events with that. In section III, we show the detection possibilities in all channels with numerical analysis result. In the last section, we give a brief conclusion of this study. 

\section{Signal and background analysis}

\begin{figure}[H]
\centering
\includegraphics[scale=0.7]{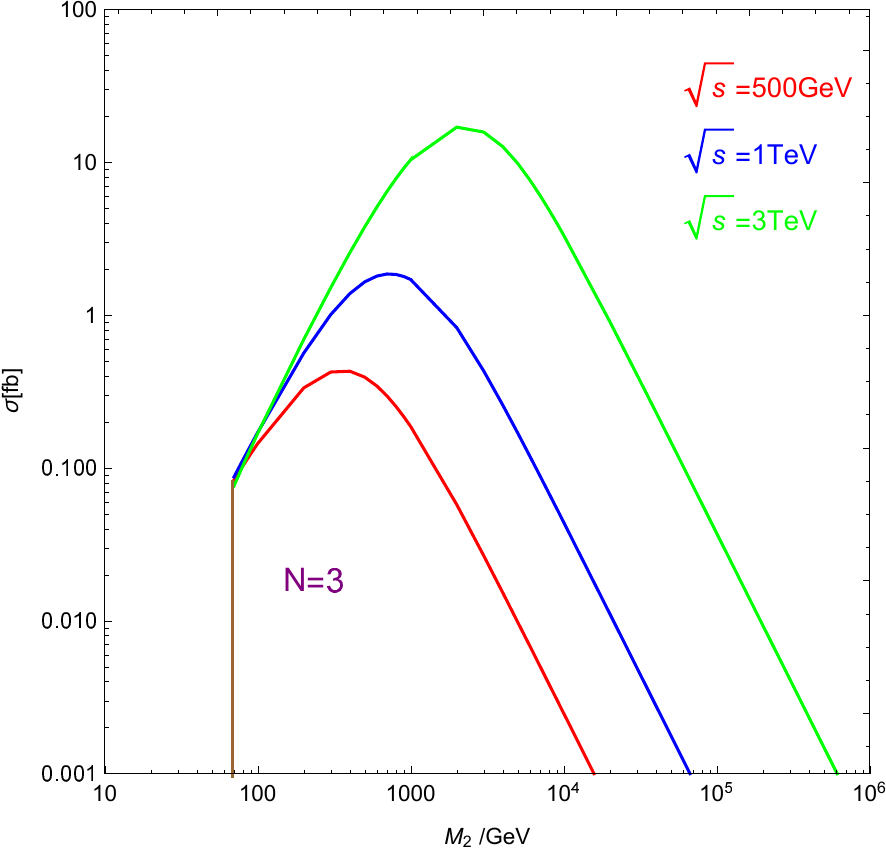}
\caption{the production cross sections of  $e^{-}e^{-}\rightarrow W^{-}W^{-}$ with $\sqrt{s}=500$~GeV~(red line), 1~TeV~(blue line) and 3~TeV~(green line)}
\label{eeww-sigma}
\end{figure}

The inverse $0\nu\beta\beta$ decay could be detected in pure leptonic, semi-leptonic and pure hadronic $W^{-}W^{-}$ decay channels. 
In Fig.\ref{eeww-sigma}, according to~\cite{Asaka:2015oia} we reproduce the cross section $\sigma(e^{-}e^{-}\to W^{-}W^{-})$ varying with $M_{2}$ in the case of three heavy Majorana neutrinos with hierarchical masses $M_{1}\ll M_{2}\ll M_{3}$.
In the following, we discuss the kinematic features of each channel and the corresponding methods to separate the signal events out from the large backgrounds.
\subsection{Pure leptonic:  $e^{-}e^{-}\to W^{-}W^{-}\rightarrow 2\ell+\slashed{E}_{T}$}

In the pure leptonic channel, the two final state leptons always move back-to-back because $W^- W^-$ is from a spin-zero system and only left-handed electrons take part in the weak interaction. This leads to a lepton angular-distribution peaking at $\cos \theta_{ll}=-1$. The $\cos \theta_{ll}$ cut could be applied to distinguish signals from backgrounds. 

On the other hand, the two invisible neutrinos make it impossible to completely reconstruct the $W$ bosons with $\slashed{E}_T$ information. $M_{T2}$ method could be used in this case by defining a minimization of all possible matches of $\slashed{{p}}_1$ and $\slashed{{p}}_2$ variables as~\cite{Lester:1999tx}
{
\bea
M_{T2}^{2}\equiv \min_{\slashed{{p}}_{1}+\slashed{{p}}_{2}=\slashed{{p}}_{T}}\left[\max\{m_{T}^{2}({p}_{T}^{\ell}, \slashed{{p}}_{1}),  m_{T}^{2}({p}_{T}^{\ell}, \slashed{{p}}_{2})\} \right]
\eea
}
where $\slashed{{p}}_T$ is the missing transverse momentum and $m_{T}$ is the reconstructed transverse mass. The $M_{T2}$ variable has an upper bound at $m_{W}$ and the corresponding $\slashed{{p}}_{1,2}$ could be used to reconstruct the system invariant mass, whose distribution is around $\sqrt{s}$ for signal events. In addition, the distinct boost effects of final-state particles should be taken into account when the collision energy is raised to several TeV, which provides us more kinematic handles on data sample reconstructions. We thus assume that the highly boosted neutrino and lepton from the same $W$ boson move approximately along the same direction. The relation $\overrightarrow{p}^{\nu}\simeq\kappa \overrightarrow{p}^{\ell}$ could now be applied and $\kappa$ is solved from 
{
\beq
\kappa = \frac{\slashed{p}_T}{\sqrt{(\overrightarrow{p}^{\ell_1}_T+\overrightarrow{p}^{\ell_2}_T)^{2}}}
\eeq
}  
Now that the four momentums of the invisible neutrinos are obtained with this approximation, the invariant-mass cut could still be applied.


\subsection{Semi-leptonic: $e^{-}e^{-}\rightarrow W^{-}W^{-}\rightarrow\ell+2j/j_{W}+\slashed{E}_{T}$}

The semi-leptonic decay has a larger signal production rate than the pure leptonic one and it's possible to reconstruct the $W^{-}W^{-}$ system. For the only missing neutrino in this symmetric collision, we can easily get its momentum with $\slashed{E}_{T}$
\bea  
\overrightarrow{p}_{\nu}=-\sum_{i}\overrightarrow{p}_{observed}.
\eea
The two on-shell $W$ bosons are then reconstructed either with a pair of jets or with the lepton and neutrino.  Similarly, when the collision energy is raised to few TeV, the boost effect becomes non-negligible and the two jets from $W^{-}$ decay would form a fat $W$-jet($j_{W}$) with its mass around $M_{W}$.


\subsection{Pure hadronic :$e^{-}e^{-}\rightarrow W^{-}W^{-}\rightarrow 4j/2j_{W}$}
In the hadronic channel with multi-jet final states, the $W$ bosons could be reconstructed with proper choices of jet-pairs and the invariant-mass of the four jets is required to be compared with $\sqrt{s}$. If the collision is energetic enough, the appearance of two $W$-jets is a key feature of this hadronic decay channel.

\subsection{Background processes }
The backgrounds of $e^{-}e^{-}\rightarrow W^{-}W^{-}$ process in different decay channels are listed in Table.\ref{bkgtable}. We would include those processes with extra electrons because of the abundance of background electrons at a $ee$-collider. These extra electrons could fake $\slashed{E}_T$ if they are not really detected, especially in the effective gauge-boson approximation and vector boson fusion processes. 
In addition, the photon radiated from the beam electron should also be considered because the cross section of backgrounds initiated from $\gamma\gamma$ collision is comparable with other channels. Its contribution is calculated in the Effective Photon Approximation with the improved Weizsaecker-Williams formula\cite{Budnev:1974de}.
\begin{table}[H]
\begin{center}
\begin{tabular}{|c|c|c|c|c|c|c|}
\hline
Process & $e^-e^-+\slashed{E}_T$ & $e^-\mu^-+\slashed{E}_T$  & $\mu^-\mu^-+\slashed{E}_T$ & $e^-+2j+\slashed{E}_T$ & $\mu^-+2j+\slashed{E}_T$ & $~~~4j~~~$\\\hline
$e^-e^-\to W^-W^-\nu_e\nu_e$ & $\bullet$ & $\bullet$ & $\bullet$ & $\bullet$ & $\bullet$ & $\bullet$ \\
$e^-e^-\to ZW^-e^-\nu_e$ & $\bullet$ & $\bullet$ & & $\bullet$ & $\bullet$ & $\bullet$ \\
$e^-e^-\to W^-e^-\nu_e$ & $\bullet$ & $\bullet$ & & $\bullet$ & & \\
$e^-e^-\to Ze^-e^-$ & $\bullet$ & & & $\bullet$ & & \\
$e^-e^-\to ZZe^-e^-$ & $\bullet$ & & & $\bullet$  & & $\bullet$ \\
$e^-e^-\to W^+W^-e^-e^-$ & & & & $\bullet$ & $\bullet$ & $\bullet$ \\
$\gamma\gamma\to W^+W^-$ & & & & $\bullet$ & $\bullet$ & $\bullet$ \\\hline
\end{tabular}
\end{center}
\caption{Backgrounds of inverse $0\nu\beta\beta$ decay process and the decay channels they contribute to}
\label{bkgtable}
\end{table}

\section{Numerical Results}
In this section, we focus on the Monte Carlo analysis of inverse $0\nu\beta\beta$ decay process. The simulation is performed with {\it MadGraph5\_v1.5.14}~\cite{Alwall:2011uj} and {\it pythia-pgs}~\cite{Sjostrand:2006za}. In order to get more kinematic features from boost effects, we choose two benchmark points with $\sqrt{s}=500$~GeV and $\sqrt{s}=3$~TeV separately. According to the previous study\cite{Asaka:2015oia}, the signal of inverse $0\nu\beta\beta$ decay with only one or two Majorana neutrino flavors are too small to be detected. For this reason, we include three heavy Majorana neutrinos in the spectrum as $M_1=3$~GeV, $M_2=350$~GeV and $M_3=35$~TeV when $\sqrt{s}=500$~GeV while $M_1=3$~GeV, $M_2=3$~TeV and $M_3=300$~TeV when $\sqrt{s}=3$~TeV. The hierarchical mass relation $M_1\ll M_2\ll M_3$ suppresses $|V_{e1}|^2$ and $|V_{e3}|^2$ to several orders smaller than $|V_{e2}|^2$ and we take the $ |V_{e}|^{2}_{\rm EW}$ value in (\ref{EWPObound}) for $|V_{e2}|^2$ accordingly. The basic cuts on final states are 
\bea
\centering
p^{\ell}_{T}>10~\text{GeV}&,&p^{j}_{T}>20~\text{GeV},\nonumber \\
|\eta^{\ell}|<2.5&,&|\eta^{j}|<5,\nonumber \\
\Delta R_{\ell\ell}>0.4&,&\Delta R_{\ell j}>0.4
\label{basiccut}
\eea
In addition, the two selected jets in the first benchmark are required to satisfy $\Delta R_{jj}>0.4$. 

\begin{figure}[H]
\centering
\includegraphics[scale=0.8]{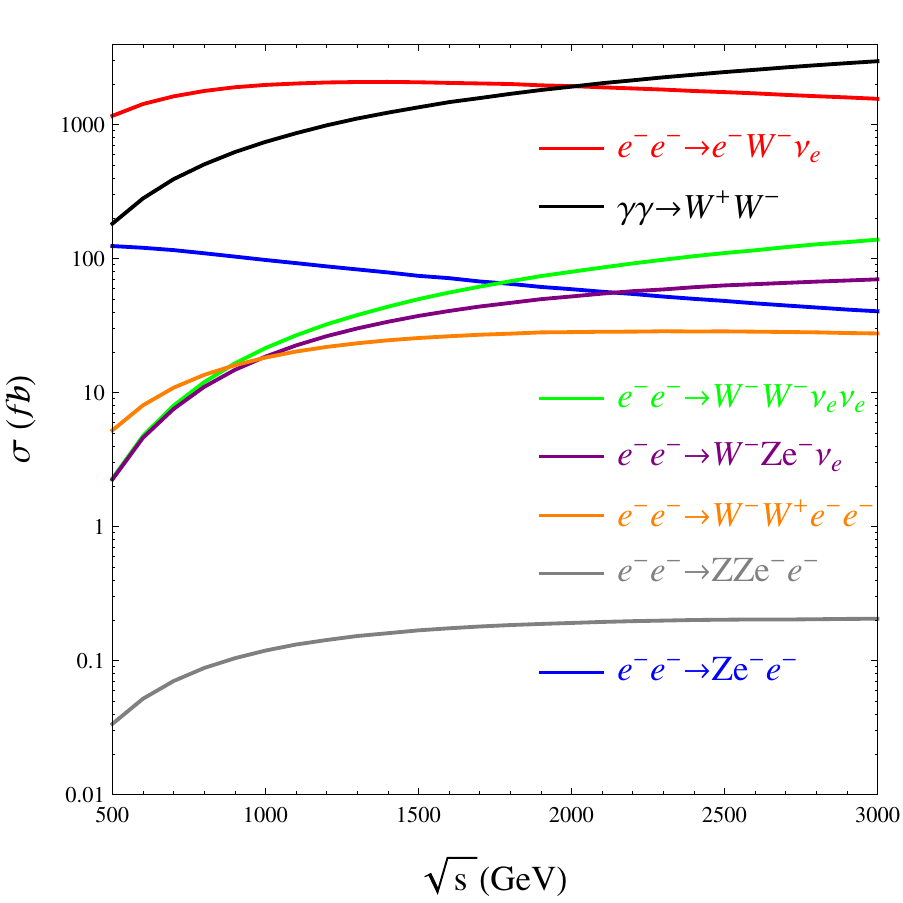}
\caption{SM background cross-sections with different $\sqrt{s}$ values}
\label{eewwbkg}
\end{figure}

In Fig.\ref{eewwbkg}, we plot SM background cross sections varying with  $\sqrt{s}$. The cross sections except for $e^-e^-\to ZZe^-e^-$ are always larger than $1~{\rm fb}$. In order to find out feasible discovery channels, we start event selection with the tagging process, which requires proper final states in each channels. For example, if more than required electrons are detected in the rapidity coverage region of the detector, they're supposed to come from background processes with extra electrons and thus we discard this event. After that, kinematic cuts are applied to eliminate background events to obtain better signal-to-background rate.

\subsection{Pure leptonic}
With $M_{T2}$ method in the first benchmark and collinear approximation in the second, we plot the reconstructed invariant-mass $m_{\rm inv}$ distributions in the $e^-e^-+\slashed{E}_T$ channel, which includes most backgrounds, in Fig.{\ref{pureleptonicfig}}a and Fig.{\ref{pureleptonicfig}}b. In order to illustrate the lepton angular correlation feature, we plot the distribution of $\cos\theta_{\ell\ell}$ in Fig.\ref{pureleptonicfig}c. We find that the signal $m_{\rm inv}$ distribution has an obviously distinguishable peak position from the backgrounds except for the $Ze^-e^-$ process.
More than that, the leptons in $Ze^-e^-$ tend more to move in the opposite directions than in other backgrounds. Alhough this $Ze^-e^-$ background has similar kinematic properties to the $e^-e^-$ signal, it should be absent in the $e^-\mu^-$ and $\mu^-\mu^-$ channels.
{
\begin{figure}[H]
\centering
\subfloat[$m_{\rm inv}$ distribution, $\sqrt{s}=500$~GeV]{\includegraphics[scale=0.350]{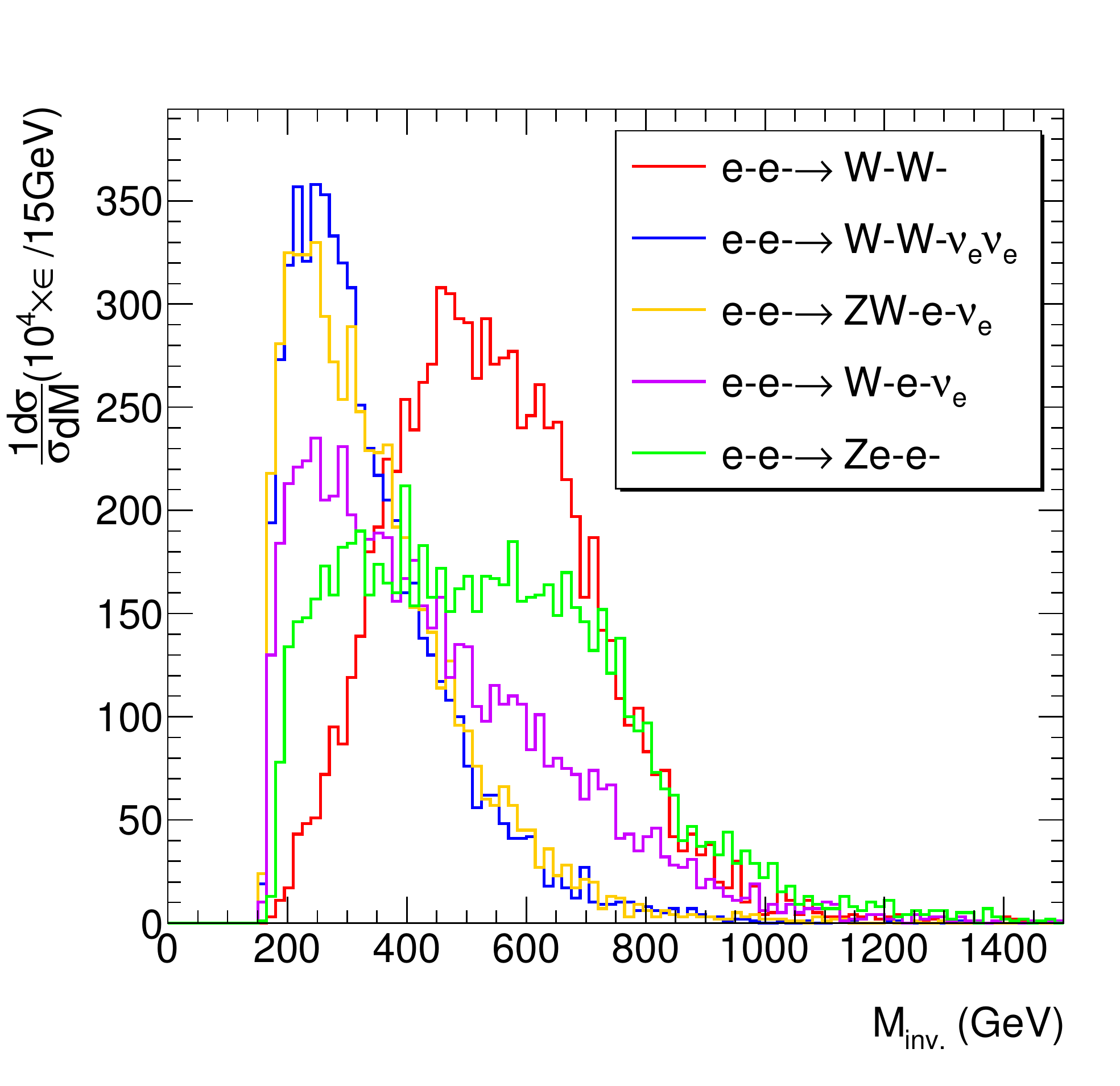}}\qquad
\subfloat[$m_{\rm inv}$ distribution, $\sqrt{s}=3$~TeV]{\includegraphics[scale=0.350]{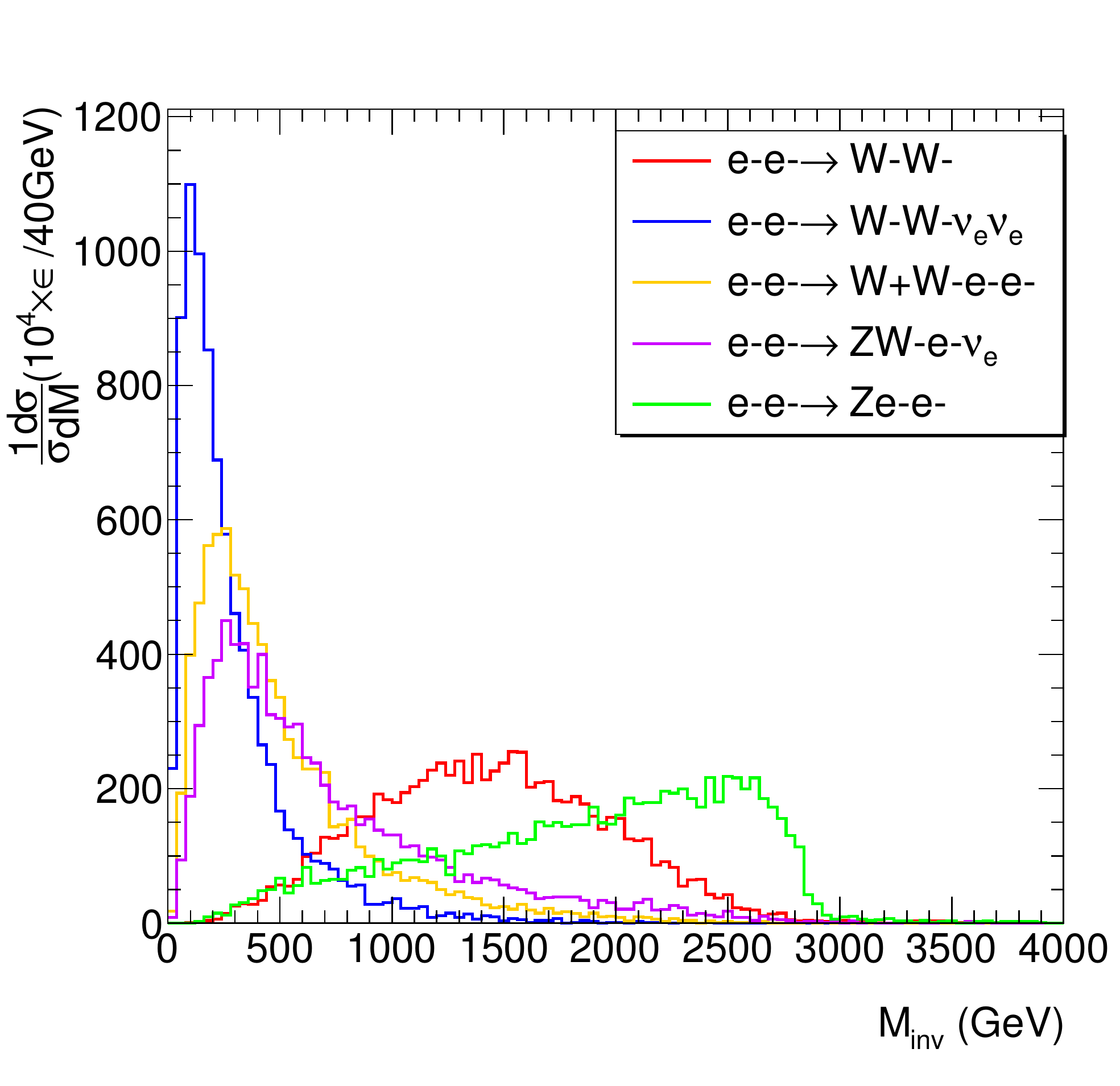}}\\
\subfloat[$\cos \theta_{ee}$ distribution, $\sqrt{s}=500$~GeV]{\includegraphics[scale=0.350]{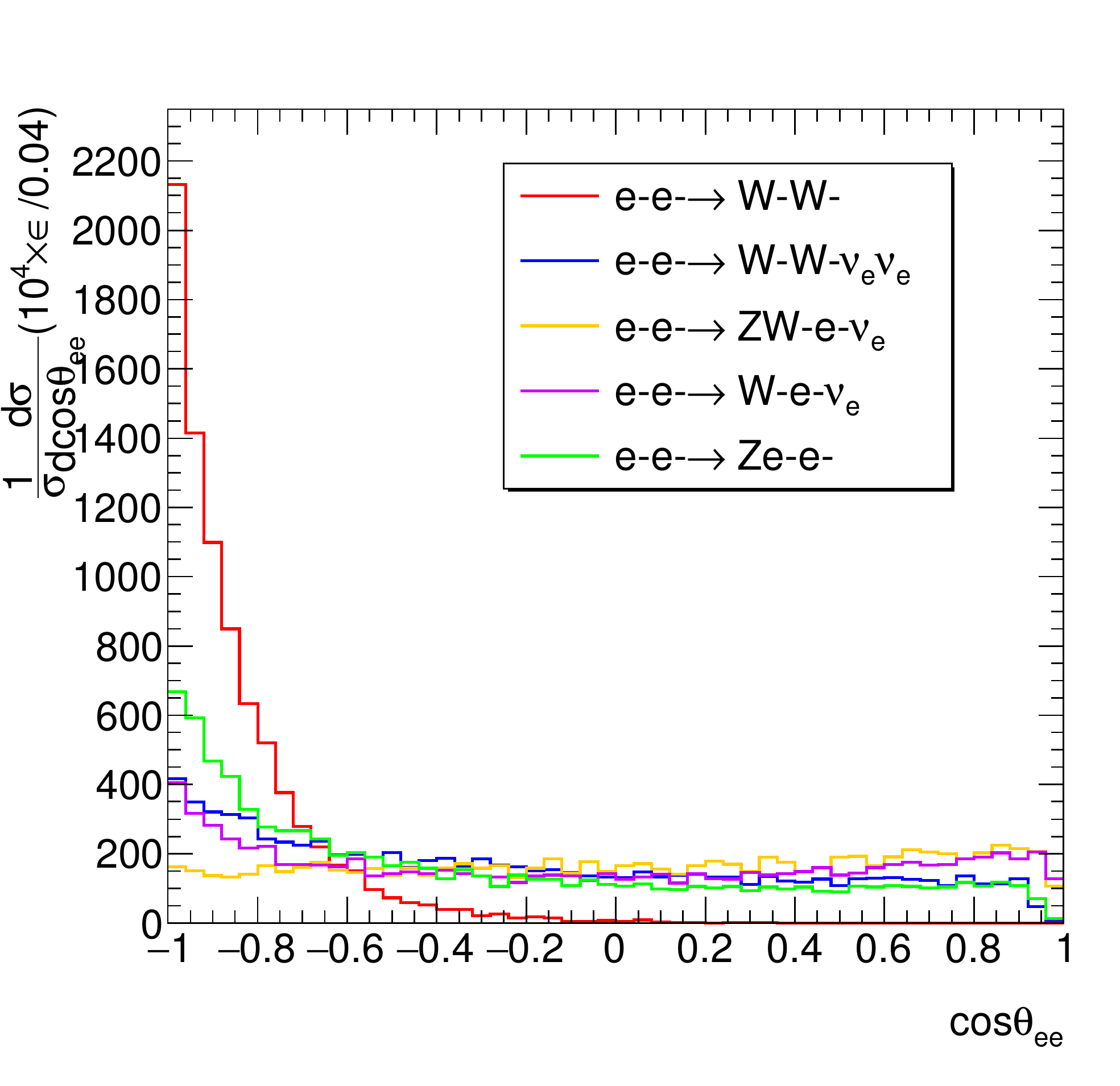}}
\caption{Kinematic features of signal and backgrounds in pure leptonic mode. $\epsilon$ is the tagging efficiency.}
\label{pureleptonicfig}
\end{figure}
}

In Table.{\ref{pureleptonic500gevtable}} and {\ref{pureleptonic3tevtable}}, we list the cross-sections after basic cuts, the survival probabilities after each kinematic cuts and the number of survived events $N$ after all cuts. ``$-$" means it's not applicable in the corresponding case. 
We assume the invariant-mass and $\cos \theta_{ll}$ cuts are independent. The $m_{\rm inv}$ cut is different in electron and muon channels to deal with different background contributions. The $\cos \theta_{\ell\ell}$ cut in the second benchmark is more severe because the signal leptons are from more boosted $W$ bosons. 
{
\begin{table}[H]
\begin{center}
\begin{tabular}{|c|c|c|c|c|c|}
 \hline
Process & $\sigma(fb)$ & $\epsilon_{\rm tagging}$  & $\epsilon_{m_{\rm inv}>400~GeV}$ & $\epsilon_{\cos\theta_{ll}<-0.7}$&$N$\\\hline
$e^-e^-+\slashed{E}_T$ channel\\\hline
$e^-e^-\to W^-W^-$ & 5.0$\times10^{-3}$ & 0.84 & 0.68 & 0.591&1\\
$e^-e^-\to W^-W^-\nu_e\nu_e $& 2.57$\times10^{-2}$ & 0.83 & 0.15 & 0.042&0\\
$e^-e^-\to ZW^-e^-\nu_e$ & 4.7$\times10^{-2}$ & 0.84 & 0.17 & 0.024&0\\
$e^-e^-\to W^-e^-\nu_e$ &120.8 & 0.83 & 0.3 & 0.069&4168\\
$e^-e^-\to Ze^-e^-$ & 24.7 & 0.84 & 0.5 & 0.185&2285\\\hline
$e^-\mu^-+\slashed{E}_T$ channel\\\hline
$e^-e^-\to W^-W^-$ & 1.0$\times10^{-2}$ & 0.87 & 0.70 & 0.603&3\\
$e^-e^-\to W^-W^-\nu_e\nu_e$ & 5.14$\times10^{-2}$ & 0.85 & 0.16 & 0.045&1\\
$e^-e^-\to ZW^-e^-\nu_e$ & 4.7$\times10^{-2}$ & 0.85 & 0.17 & 0.024&0\\
$e^-e^-\to W^-e^-\nu_e$ & 120.8 & 0.80 & 0.29 & 0.069&4168\\\hline
$\mu^-\mu^-+\slashed{E}_T$ channel\\\hline
$e^-e^-\to W^-W^-$ & 5.0$\times10^{-3}$ & 0.90 & 0.73 & 0.633&1\\
$e^-e^-\to W^-W^- \nu_e\nu_e$ & 2.57$\times10^{-2}$ & 0.87 & 0.16 & 0.044&0\\\hline
\end{tabular}
\end{center}
\caption{Cross-section and cut efficiencies in pure leptonic mode with $\sqrt{s}=500$~GeV} and $\mathcal{L}=500~$fb$^{-1}$
\label{pureleptonic500gevtable}
\end{table}
}

{
\begin{table}[H]
\begin{center}
\begin{tabular}{|c|c|c|c|c|c|c|c|}
\hline
Process & $\sigma(fb)$ & $\epsilon_{\rm tagging}$  & $\epsilon_{0.9~{\rm TeV}<m_{\rm inv}<1.9~{\rm TeV}}$ & $\epsilon_{m_{\rm inv}>900~{\rm GeV}}$ & $\epsilon_{m_{\rm inv} > 700~GeV}$ & $\epsilon_{\cos\theta_{ll}<-0.95}$ &$N$\\\hline
$e^-e^-+\slashed{E}_T$ channel\\\hline
$e^-e^-\to W^-W^-$ & 0.18 & 0.82 & 0.52 & $-$ & $-$ & 0.52 &47\\
$e^-e^-\to W^-W^- \nu_e\nu_e$ & 1.3 & 0.83 & 0.03 & $-$ & $-$ & 0.0018 &1\\
$e^-e^-\to ZW^-e^-\nu_e$ & 1.15& 0.83 & 0.1 & $-$ & $-$ & 0.0024 &1\\
$e^-e^-\to W^-e^-\nu_e$ &124.5& 0.83 & 0.18 & $-$ & $-$ & 0.0173 &1077\\
$e^-e^-\to Ze^-e^-$ & 8 & 0.82 & 0.29 & $-$ & $-$ & 0.152 &608\\\hline
$e^-\mu^-+\slashed{E}_T$ channel\\\hline
$e^-e^-\to W^-W^-$ & 0.37 & 0.86 & $-$ & 0.72 & $-$ & 0.72 &133\\
$e^-e^-\to W^-W^- \nu_e\nu_{e}$ & 2.6& 0.83 & $-$ & 0.03 & $-$ & 0.0018 &2\\
$e^-e^-\to ZW^-e^-\nu_e$ & 1.15 & 0.82 & $-$ & 0.11 & $-$ & 0.0027 &2\\
$e^-e^-\to W^-e^-\nu_e$ & 124.5 & 0.83 & $-$ & 0.23 & $-$ & 0.0222 &1382\\\hline
$\mu^-\mu^-+\slashed{E}_T$ channel\\\hline
$e^-e^-\to W^-W^-$ & 0.18 & 0.90 & $-$ & $-$ & 0.83 & 0.8208 &74\\
$e^-e^-\to W^-W^- \nu_e\nu_e$ & 1.3 & 0.82 & $-$ & $-$ & 0.06 & 0.0037 &2\\\hline
\end{tabular}
\end{center}
\caption{Cross-section and cut efficiencies in pure leptonic mode with $\sqrt{s}=3$~TeV and $\mathcal{L}=500~$fb$^{-1}$}
\label{pureleptonic3tevtable}
\end{table}
}

\subsection{Semi-leptonic}
\begin{figure}[H]
\centering
\subfloat[$m_{\rm inv}$ distribution, $\sqrt{s}=500$~GeV]{\includegraphics[scale=0.350]{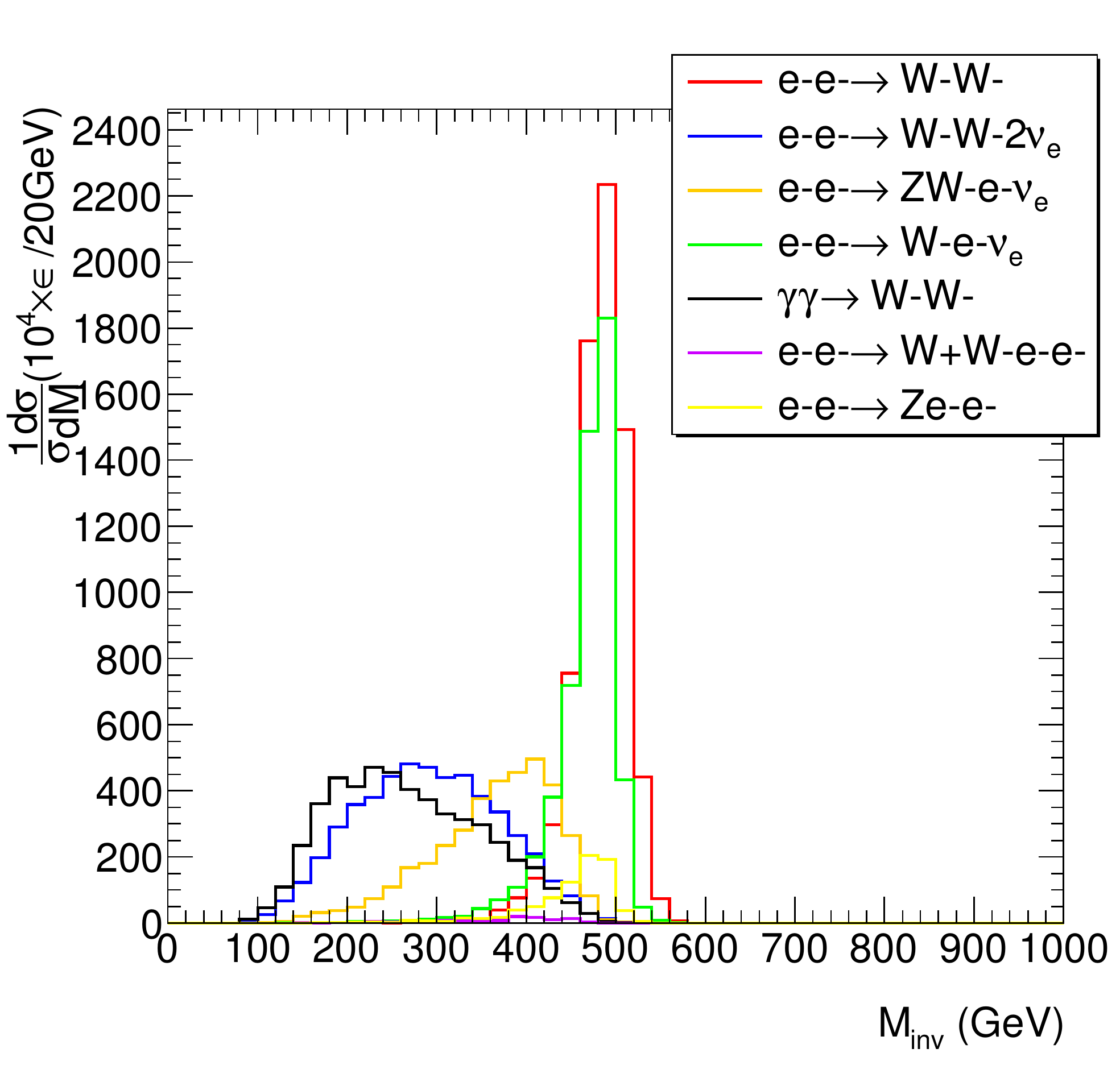}}\qquad
\subfloat[$m_{W}$ distribution, $\sqrt{s}=500$~GeV]{\includegraphics[scale=0.350]{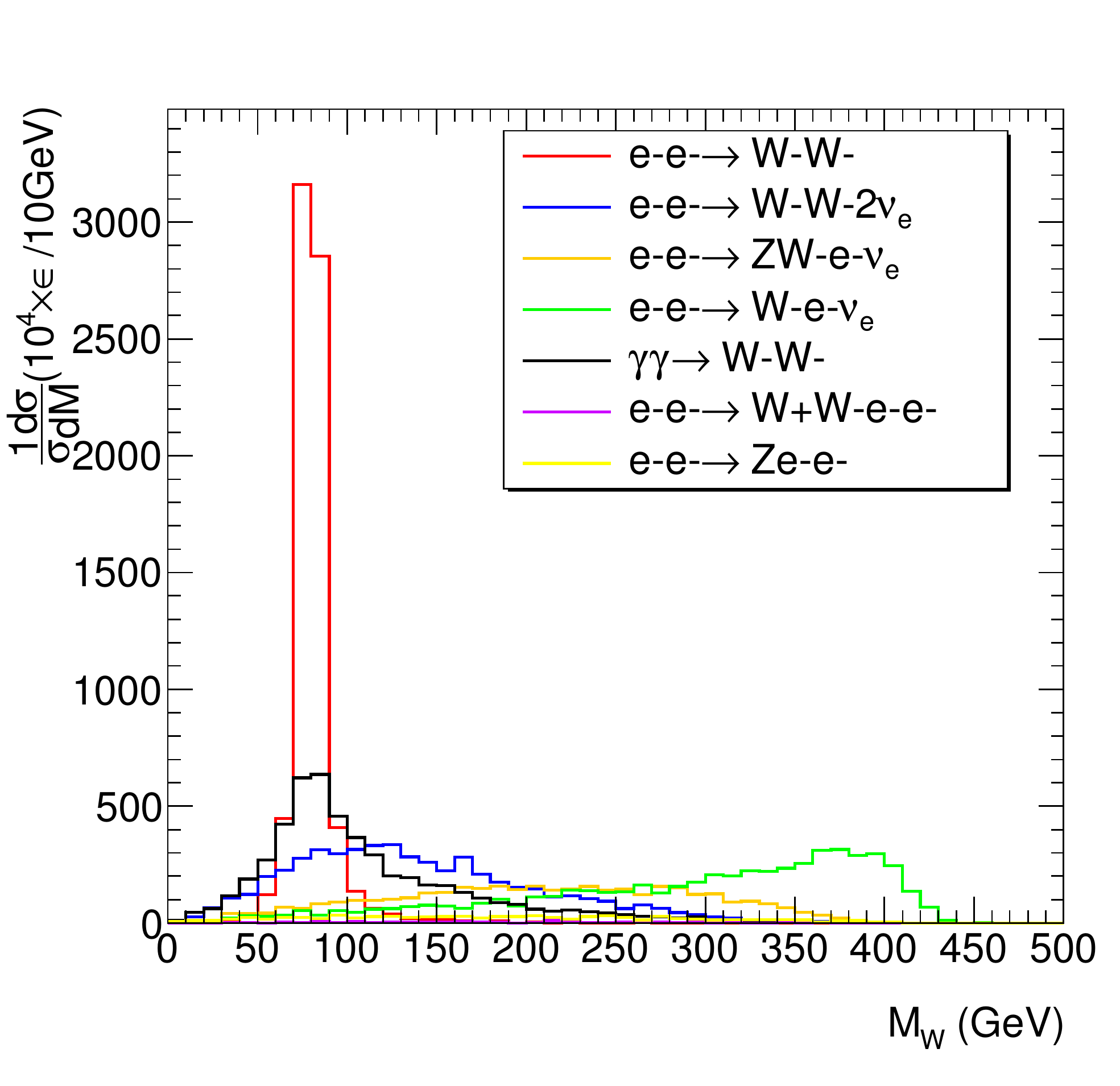}}\\
\caption{$m_{\rm inv}$ and $m_{W}$ distributions of the reconstructed system in $e^-+2j+\slashed{E}_T$ channel with $\sqrt{s}=500$~GeV. $\epsilon$ is the tagging efficiency.}
\label{semileptonicfig}
\end{figure}

The semi-leptonic channel can be completely reconstructed because there's only one invisible neutrino in the final states. The system reconstructed with $\slashed{E}_T$ and lepton is identified as a $W$ boson, whose mass distribution could be used to cut out $W^-e^-\nu_e$ background in the $e^-+2j+\slashed{E}_T$ channel. The $\gamma \gamma \to W^+ W^-$ also contains reconstructable $W$-pair, but we are to use an invariant-mass cut to suppress it. In Fig.\ref{semileptonicfig}a and Fig.\ref{semileptonicfig}b, the $m_{\rm inv}$ and ${m_W}$ distributions in $e^-+2j+\slashed{E}_T$ channel are presented. The distributions of $\gamma \gamma \to W^+ W^-$ and $e^{-}e^{-}\to W^-e^-\nu_{e}$ processes can mimic the signal $m_{W}$ and $m_{\rm inv}$ distributions respectively, but not both. In the $3$~TeV case, the hadronic $W$ is identified as $j_{W}$ according to the discussion in the last section.

The $M_{W}$ and $M_{\rm inv}$ cuts are powerful in signal event selection and we list the  survival efficiencies and event numbers after successive cuts in Table.{\ref{semileptonic500gevtable}} and {\ref{semileptonic3tevtable}}.
{
\begin{table}[H]
\begin{center}
\begin{tabular}{|c|c|c|c|c|c|c|}
 \hline
Process & $\sigma(fb)$& $\epsilon_{\rm tagging}$  & $\epsilon_{400~GeV<m_{\rm inv}<550~GeV}$ & $\epsilon_{m_{\rm inv}>400~GeV}$ & $\epsilon_{70<m_{W}< 90~GeV}$&$N$ \\\hline
$e^-+2j+\slashed{E}_T$ channel \\\hline
$e^-e^-\to W^-W^-$ & 5.64$\times 10^{-2}$ & 0.74 & 0.72 & $-$ & 0.6&17\\
$e^-e^-\to W^-W^-\nu_e\nu_e$ & 0.23 & 0.52 & 0.046 & $-$ & 0.003&0\\
$e^-e^-\to ZW^-e^-\nu_e$ & 0.3 &0.37 &0.13& $-$ &0.002&0 \\
$e^-e^-\to W^-e^-\nu_e$ & 537.3 & 0.54 & 0.51 & $-$ & 0.007&1880\\
$e^-e^-\to W^+W^-e^-e^-$  & 0.23 & 0.01 & 0.004 & $-$ & 0.0002&0\\
$e^-e^-\to Ze^-e^-$ & 49.1 & 0.08 & 0.07 & $-$ & 0.003&74\\
$\gamma\gamma\to W^+W^-$ & 8$~{\rm fb}$ & 0.51 & 0.037 & $-$ & 0.006&24\\\hline
$\mu^-+2j+\slashed{E}_T$ channel\\\hline
$e^-e^-\to W^-W^-$ & 5.64$\times10^{-2}$  & 0.74 & $-$ & 0.72 & 0.6&17\\
$e^-e^-\to W^-W^-\nu_e\nu_e $ & 0.23 & 0.52 & $-$ & 0.05 & 0.002&0\\
$e^-e^-\to ZW^-e^-\nu_e $ & 0.1 &0.04& $-$ &0.01&0.0004&0\\
$e^-e^-\to W^+W^-e^-e^-$ & 0.23 & 0.037 & $-$ & 0.0008 & 0.0001&0\\
$\gamma\gamma\to W^+W^-$ & 8 & 0.49 & $-$ & 0.04 & 0.003&12\\\hline
\end{tabular}
\end{center}
\caption{Cross-section and cut efficiencies in semi-leptonic channel with $\sqrt{s}=500$~GeV and $\mathcal{L}=500~$fb$^{-1}$}
\label{semileptonic500gevtable}
\end{table}
}

{
\begin{table}[H]
\begin{center}
\begin{tabular}{|c|c|c|c|c|}
 \hline
Process & $\sigma(fb)$ & $\epsilon_{\rm tagging}$  & $\epsilon_{m_{\rm inv}>2.5~TeV}$&$N$  \\\hline
$e^-+j_{W}+\slashed{E}_T$ channel \\\hline
$e^-e^-\to W^-W^-$ & 2.2 & 0.78 & 0.77& 847 \\
$e^-e^-\to W^-W^-\nu_e\nu_e $ & 13.2 & 0.062 & 0.0032& 21 \\
$e^-e^-\to ZW^-e^-\nu_e$ & 9.1 &0.065  &0.0064& 29  \\
$e^-e^-\to W^-e^-\nu_e$ & 774.5 & 0.098 & 0.018& 6970 \\
$e^-e^-\to W^+W^-e^-e^-$ &1.143 & 0.0013 & 0.0003& 0 \\
$e^-e^-\to Ze^-e^-$ & 15.76 & 0.008 & $<0.0001$&  0 \\
$\gamma\gamma\to W^+W^-$ & 113 & 0.006 & 0.0003&17   \\\hline
$\mu^-+j_{W}+\slashed{E}_T$ channel\\\hline
$e^-e^-\to W^-W^-$ & 2.2 & 0.75 & 0.75&825  \\
$e^-e^-\to W^-W^-\nu_e\nu_e$ & 13.2 & 0.06 & 0.0026& 17 \\
$e^-e^-\to ZW^-e^-\nu_e$ & 2.4 &0.0034 &$<0.0001$&0  \\
$e^-e^-\to W^+W^-e^-e^-$ & 1.143 & 0.0009 & 0.0001 & 0\\
$\gamma\gamma\to W^+W^-$ & 113 & 0.005 & 0.0002 & 11\\\hline
\end{tabular}
\end{center}
\caption{Cross-section and cut efficiencies in  semi-leptonic channel with $\sqrt{s}=3$~TeV and $\mathcal{L}=500~$fb$^{-1}$}
\label{semileptonic3tevtable}
\end{table}
}

\subsection{Pure hadronic }
In the hadronic decay channel, the four jets are chosen to reconstruct the complete system. The invariant-mass distributions of each processes are shown in Fig.\ref{hadronicfig}. It is clear that $m_{\rm 4j}$ distributions of the backgrounds deviate significantly from $\sqrt{s}$ either because there're undetected leptons carrying away part of the energy or because the process is a photon-photon scattering.
{
\begin{figure}[H]
\centering
{\includegraphics[scale=0.350]{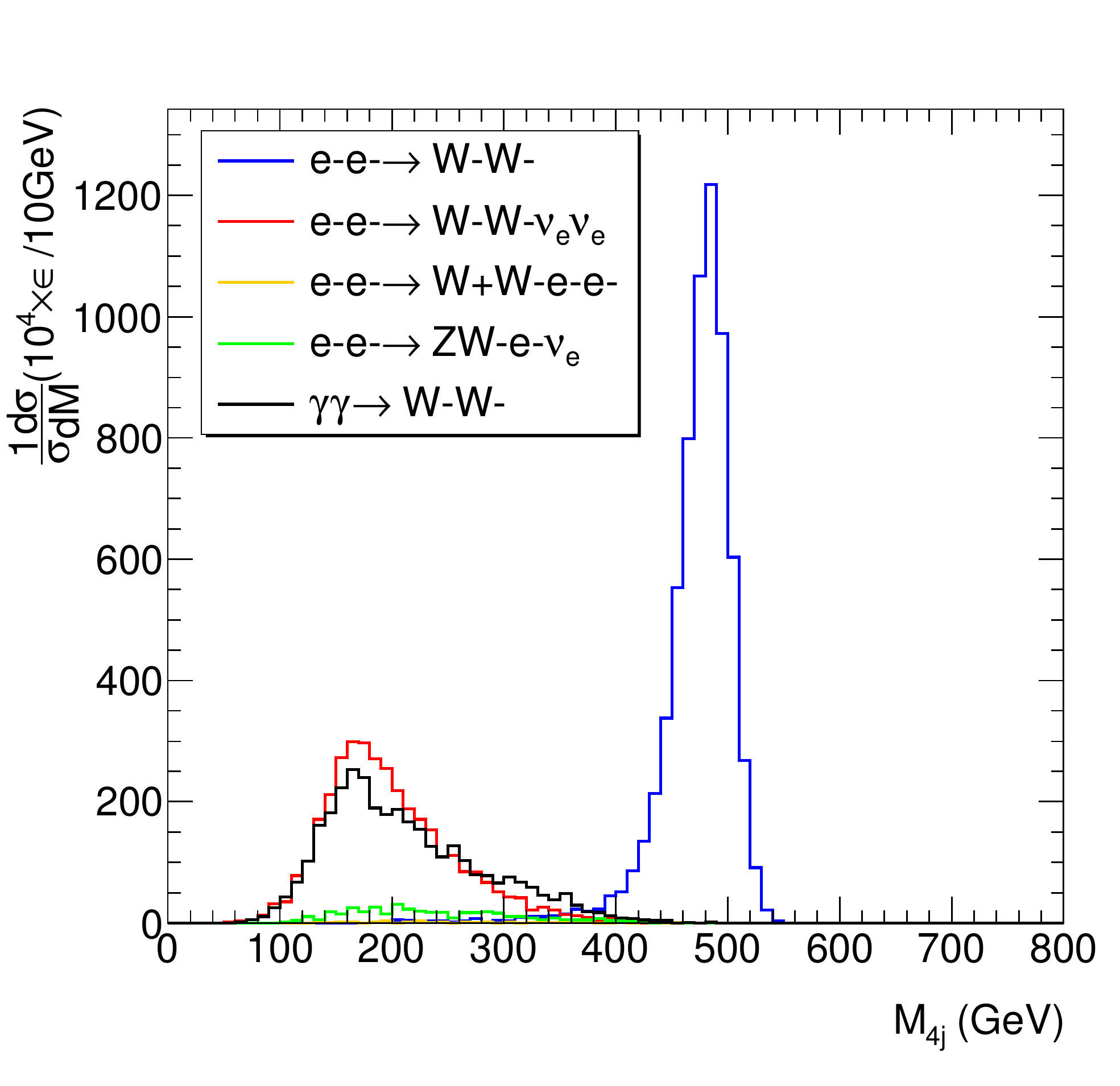}}
\caption{$m_{\rm 4j}$ distribution of the four-jet system, $\sqrt{s}=500~$GeV. $\epsilon$ is the tagging efficiencies}
\label{hadronicfig}
\end{figure}
}
In the $\sqrt{s}=3$~TeV case, the tagging process requires two highly boosted $W$-jets with jet mass around $m_{W}$ and cone size small enough. We further require the separation between two $W$-jets be larger than $0.4$. The gauge bosons from the backgrounds are not that boosted since electrons and neutrinos in final states carry away large energy. This is also true for $\gamma\gamma$ process because radiated photons are not so energetic as the electrons. We find the cone size values, which could be estimated with the separations between $W$ hadronic decay final states, are in general larger in background events. Thus the background events can hardly meet the $j_{W}$ tagging criteria. The detailed survival efficiencies and number of events after implementing all cuts are listed in the tables below.

{
\begin{table}[H]
\begin{center}
\begin{tabular}{|c|c|c|c|c|}
 \hline
Process & $\sigma(fb)$ & $\epsilon_{\rm tagging}$  & $\epsilon_{m_{\rm 4j}>400~GeV}$ &$N$\\\hline
$4j$ channel \\\hline
$e^-e^-\to W^-W^-$ & 0.16 & 0.66 & 0.64&51\\
$e^-e^-\to W^-W^-\nu_e\nu_e$ & 0.5 & 0.35 & 0.0006&0\\
$e^-e^-\to W^+W^-e^-e^-$ & 1 & 0.004 & 0.0005&0\\
$e^-e^-\to ZW^-e^-\nu_e $ & 0.4 & 0.04 & 0.0008&0\\
$\gamma\gamma\to W^+W^- $ & 34.4& 0.33 & 0.0031&53\\\hline
\end{tabular}
\end{center}
\caption{Cross-section and cut efficiencies in  hadronic channel with $\sqrt{s}=500$~GeV and $\mathcal{L}=500~$fb$^{-1}$}
\label{hadronic500gevtable}
\end{table}
}
{
\begin{table}[H]
\begin{center}
\begin{tabular}{|c|c|c|c|c|}
 \hline
Process & $\sigma(fb)$ & $\epsilon_{\rm tagging}$  & $\epsilon_{m_{\rm 4j}>2.3~TeV}$&$N$ \\\hline
$2j_{W}$ channel\\\hline
$e^-e^-\to W^-W^-$ & 6.7& 0.73 & 0.73&2446\\
$e^-e^-\to W^-W^-\nu_e\nu_e$ & 34.4 & 0.011 & 0.0001&2\\
$e^-e^-\to W^+W^-e^-e^-$ & 6.3 & 0.0001 & $<0.0001$&0\\
$e^-e^-\to ZW^-e^-\nu_e$ & 14.3 & 0.0006 & $<0.0001$&0\\
$\gamma\gamma\to W^+W^-$ & 602 & 0.0033 & $<0.0001$ &30\\\hline
\end{tabular}
\end{center}
\caption{Cross-section and cut efficiencies in  hadronic channel with $\sqrt{s}=3$~TeV and $\mathcal{L}=500~$fb$^{-1}$}
\label{hadronic3tevtable}
\end{table}
}

\subsection{Detection possibility}
At last, we use the signal-to-bakcground ratio $\frac{S}{B}$ and significance $s=\frac{S}{\sqrt{S+B}}$ to evaluate the detection possibility in each channel with $\mathcal{L}=500~$fb$^{-1}$. The channels in which inverse $0\nu\beta\beta$ decay could be detected are listed in Table.\ref{significancetable}. For $\sqrt{s}=3~$TeV the channels with large signal-to-background ratio, the approximate expression for $s$ is not valid but we argue the detection could be through event counting. The pure hadronic channel with $\sqrt{s}=500~$GeV and for $\sqrt{s}=3~$TeV the semi-leptonic channel with electron are also viable for inverse $0\nu\beta\beta$ decay detection with $5\sigma$ significance. The $\sqrt{s}=500~$GeV semi-leptonic channel with muon and $\sqrt{s}=3~$TeV pure leptonic channel with $e^{-}\mu^{-}$ still require $750$ fb$^{-1}$ and $1000$ fb$^{-1}$ integrated luminosity respectively for a detection.

{
\begin{table}[H]
\begin{center}
\begin{tabular}{|c|c|c|c|c|}
\hline
 \multirow{2}{*}{Process}&\multicolumn{2}{|c|}{$\sqrt{s}=500$~GeV} &\multicolumn{2}{|c|}{$ \sqrt{s}=3$~TeV} \\
 \cline{2-5}
&$\frac{S}{B}$&$s$&$\frac{S}{B}$&$s$\\\hline
$e^-\mu^-+\slashed{E}_T$&$-$&$-$ &0.1 &3.4 \\
$\mu^-\mu^-+\slashed{E}_T$ &$-$&$-$&37.0 &8.5 \\
$e^-+ 2j/j_{W}+\slashed{E}_T$&$-$&$-$  &0.1&9.5 \\
$\mu^-+2j/j_{W}+\slashed{E}_T$ &1.4&3.1 &29.5&28.2\\
$4j/2j_{W}$ & 0.95&5.0&76.4 &49.1
 \\\hline
\end{tabular}
\end{center}
\caption{Signal-to-background ratio and  Significance in different decay channels with $\mathcal{L}=500~$fb$^{-1}$}
\label{significancetable}
\end{table}
}

\begin{figure}[H]
\centering
\includegraphics[scale=0.8]{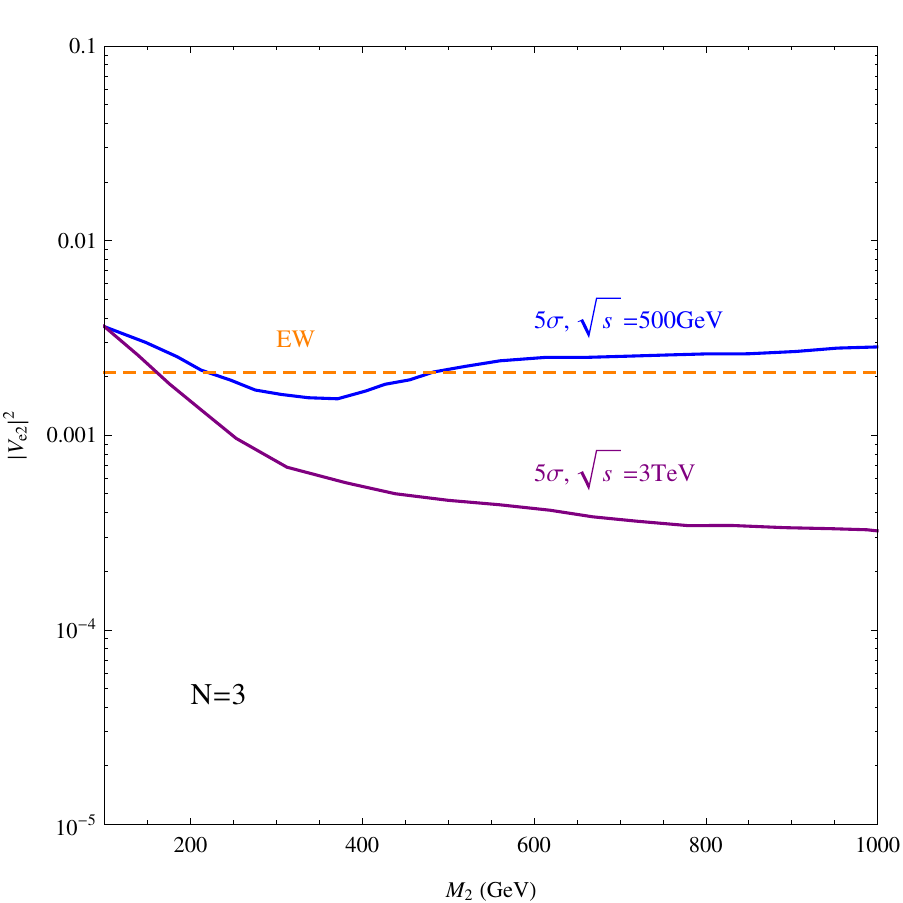}
\caption{$5\sigma$ exclusion limit of $|V_{e2}|^{2}$ with varying $M_{2}$ in pure hadronic channel}
\label{signal-5sigma-4j}
\end{figure}

In Fig.\ref{signal-5sigma-4j}, we present a comparison between $|V_{e2}|^{2}$ exclusion limit in the pure hadronic decay mode with $\mathcal{L}=500~{\rm fb}^{-1}$ and the EWPO bound. 
We find the $\sqrt{s}=500~$GeV option has only limited advantage over current bound in the region $250~{\rm GeV}\lesssim M_{2}\lesssim 450~{\rm GeV}$. But for the $\sqrt{s}=3~$TeV case, the exclusion limit on $|V_{e2}|^{2}$ could reach $\mathcal{O}(10^{-4})$ when $M_{2}\gtrsim150~$GeV, providing a chance to probe Majorana neutrinos beyond EWPO experiments.

\section{Conclusion}

The $e^{-}e^{-}\to W^{-}W^{-}$ scattering may potentially become an important realization of $0\nu\beta\beta$ decay at future electron colliders, which provides an alternative way to probe the Majorana nature of neutrinos. There're several advantages of inverse $0\nu\beta\beta$ decay search. First of all, this process is free from the nuclear matrix element uncertainties. Secondly, due to the difference of energy scale from double-beta decays, 
the $e^{-}e^{-}\to W^{-}W^{-}$ scattering may become a complementary test to probe the LNV processes, particularly in the parameter region where significant destructive interference occurs in double-beta decays. 
In this study, we focus on collider phenomenology of $e^{-}e^{-}\to W^{-}W^{-}$ process and find the kinematic features that help to increase the detection potential. For example, the $M_{T2}$ method and lepton angular distribution $\theta_{\ell\ell}$ are quite effective in the pure leptonic channel. The boost effects in the $\sqrt{s}=3~$TeV case allow us to apply $j_W$ tagging and the collinear approximation for $W$ decay products. We get better numerical analysis result in the pure hadronic channel and those with $W$ decaying leptonically to muon, while the abundant electron background's influence on $e^{-}e^{-}+\slashed{E}_{T}$ and $e^{-}+2j+\slashed{E}_{T}$ channels is not a negligible issue. We then translate the results into signal-to-background ratio and significance to evaluate detection possibility. In the $\sqrt{s}=500~$GeV case with $\mathcal{L}=500~$fb$^{-1}$, the pure hadronic channel could already provide a $5\sigma$ detection. If we raise the collision energy to $3~$TeV, the inverse $0\nu\beta\beta$ decay process could be detected in pure hadronic channel, semi-leptonic channel with muon and pure leptonic channel with dimuon simply through event counting. And if $1000$fb$^{-1}$ data are available, $5\sigma$ detection could also be made in both $500~$GeV semi-leptonic channel with muon and $3~$TeV pure leptonic channel with $e^{-}\mu^{-}$. The pure hadronic channel result is used to constrain heavy neutrino mixing in the $|V_{e2}|^{2}$-$M_{2}$ plane. The result shows that the $500~$GeV c.m. energy exclusion is weaker than current EWPO bound except for a small region around $350~$GeV while the $\sqrt{s}=3$~TeV exclusion limit is significantly stronger, reaching $\mathcal{O}(10^{-4})$. This indicates the important role of inverse $0\nu\beta\beta$ decay in future Majorana neutrino searches, especially at a electron collider with higher energy and luminosity.

\section{Acknowledgement}

The work is supported in part by the National Science Foundation of China (11135006,  11275168, 11422544, 11075139, 11375151, 11535002) and the Zhejiang University Fundamental Research Funds for the Central Universities. KW is also supported by Zhejiang University K.P.Chao High Technology Development Foundation.

\end{document}